\documentclass[preprint,12pt]{elsarticle}

\usepackage{framed,multirow}
\usepackage{amsmath,amssymb,amsfonts,amsthm}
\usepackage{latexsym}
\usepackage{url}
\usepackage{xcolor}
\usepackage{hyperref}
\usepackage{algorithmic}
\usepackage{graphicx,epstopdf}
\usepackage{textcomp}
\usepackage{mathrsfs}
\usepackage{tensor}
\usepackage{esint}
\usepackage{tikz}
\usepackage{multirow}
\usepackage{tabularx}
\usepackage{adjustbox}

\usepackage{makecell}
\usepackage{textcomp}

\journal{Arxiv}

\newcolumntype{C}[1]{>{\centering\arraybackslash}m{#1}}

\newcommand{\x}{\mbox{\textbf{x}}}

\DeclareMathOperator*{\argmin}{\arg\min}
\DeclareMathOperator*{\argmax}{\arg\max}

                    {$\blacksquare$\vspace*{7pt}} % square <--> blacksquare

\def\f{\frac}

\def\bi{{\mathbf i}}

\def\h{{\mathbf h}}
\def\n{\mathbf{n}}

\def\w{{\boldsymbol w}}

\def\c{\boldsymbol{c}}
\def\x{\boldsymbol{x}}

\def\z{{\boldsymbol z}}

\def\R{{\mathbb R}}

\def\mL{{\mathcal L}}

\def\mS{{\mathcal S}}

\def\bi{\begin{itemize}} \def\ei{\end{itemize}}
\def\be{\begin{eqnarray*}}
\def\ee{\end{eqnarray*}}

\def\etal{{\it et al }}

\def\0{{\mathbf 0}}

\newcommand{\beq}{\begin{equation}}
\newcommand{\eeq}{\end{equation}}

\def\eref#1{(\ref{#1})}

\newcommand{\eps}{\varepsilon}

\def\x{\mathbf{x}}

\def\z{\mathbf{z}}

\def\XXint#1#2#3{{\setbox0=\hbox{$#1{#2#3}{\int}$ }
\vcenter{\hbox{$#2#3$ }}\kern-.55\wd0}}

\begin{document}

%% \verso{xxx \textit{et~al.}}

\begin{frontmatter}
\title{A robust multi-domain network for short-scanning amyloid PET reconstruction}

\author[1]{Hyoung Suk Park \fnref{*}}
\author[2]{Young Jin Jeong \fnref{*}}
\author[1]{Kiwan Jeon\corref{cor1}}
\cortext[cor1]{Corresponding author:\,\, Kiwan Jeon~(jeonkiwan@nims.re.kr)}
\author[]{for the Alzheimer's Disease Neuroimagin Initiative \fnref{fn1}}

\fntext[*]{These authors contributed equally to this work.}
\fntext[fn1]{Data used in preparation of this article were obtained from the Alzheimer's Disease Neuroimaging Initiative (ADNI) database (adni.loni.usc.edu). As such, the investigators within the ADNI contributed to the design and implementation of ADNI and/or provided data but did not participate in analysis or writing of this report. A complete listing of ADNI investigators can be found at: to the design and implementation of ADNI and/or provided data but did not participate in analysis or writing of this report. A complete listing of ADNI investigators can be found at:\url{http://adni.loni.usc.edu/wp-content/uploads/how_to_apply/ADNI_Acknowledgement_List.pdf}}

\address[1]{National Institute for Mathematical Sciences, Daejeon, 34047, Republic of Korea}
\address[2]{Department of Nuclear Medicine, Dong-A University Hospital and College of Medicine, Dong-A University, Busan, 49201, Republic of Korea}

\begin{abstract}
This paper presents a robust multi-domain network designed to restore low-quality amyloid PET images acquired in a short period of time. The proposed method is trained on pairs of PET images from short (2 minutes) and standard (20 minutes) scanning times, sourced from multiple domains. Learning relevant image features between these domains with a single network is challenging. Our key contribution is the introduction of a mapping label, which enables effective learning of specific representations between different domains. The network, trained with various mapping labels, can efficiently correct amyloid PET datasets in multiple training domains and unseen domains, such as those obtained with new radiotracers, acquisition protocols, or PET scanners. Internal, temporal, and external validations demonstrate the effectiveness of the proposed method. Notably, for external validation datasets from unseen domains, the proposed method achieved comparable or superior results relative to methods trained with these datasets, in terms of quantitative metrics such as normalized root mean-square error and structure similarity index measure. Two nuclear medicine physicians evaluated the amyloid status as positive or negative for the external validation datasets, with accuracies of 0.970 and 0.930 for readers 1 and 2, respectively. 
\end{abstract}

\end{frontmatter}

\section{Introduction}
Amyloid positron emission tomography (PET) is a widely utilized nuclear medicine imaging modality for detecting amyloid plaques in patients with memory disorders, such as Alzheimer's disease. Various radiotracers, including F-18 florbetaben (FBB), F-18 florapronol (FPN), F-18 flutemetamol (FMM), and F-18 florbetapir (FBP), are commercially available for amyloid PET imaging, typically requiring a scanning time of 20-30 minutes. However, elderly patients with memory disorders often struggle to remain still for such an extended period, resulting in head movements that can cause motion artifacts in PET images and degrade their diagnostic value. In some instances, patients may require re-scanning or additional radiation exposure due to poor image quality caused by movement. Consequently, there is an increasing demand for reducing scan time while maintaining diagnostic reliability. Nonetheless, PET images acquired from shorter scanning times can suffer from low signal-to-noise ratios and diminished diagnostic reliability, potentially limiting their clinical utility.

In recent years, deep learning techniques have demonstrated significant potential for improving PET image quality by reducing noise \cite{Arabi2021,Wang2020,Zhou2021}. Deep learning models learn the relationship between low-quality and high-quality PET images obtained from source and target domains. Several studies have focused on minimizing noise caused by reduced radiotracer injection using various deep learning techniques, such as convolutional neural network (CNN) \cite{Chen2019,Gong2018,Wang2021} and generative adversarial networks (GAN) \cite{Ouyang2019,Xiang2017,Zhou2020}. Some research has attempted to decrease noise resulting from shortened PET acquisition times and enhance image quality. For example, Jeong \etal. \cite{Jeong2021} and Sanaat \etal. \cite{Sanaat2021} employed deep learning for short-scanning PET image restoration. They used multiple pairs of PET images acquired at short-scanning times of 2 and 3 minutes and standard scanning times of 20 and 27 minutes, respectively, for training purposes.

Nonetheless, most existing methods have been trained and validated using PET images from a single source domain. PET image domains can exhibit different characteristics depending on various factors, such as the type of radiotracer, PET scanner, reconstruction method, and scanning protocol. Due to these domain discrepancies, the correction performance of deep learning methods may significantly degrade when applied to PET images from an unseen source domain.

In this paper, we propose a deep learning-based method for short-scanning amyloid PET image restoration that offers enhanced generalization across various image domains. To improve generalization, we simultaneously learn a correction map using amyloid PET images sampled from multiple source and target domains. However, capturing all relevant image features between source and target domains with a single correction network is challenging \cite{Joshi2012,Nam2016}. To address this issue, we introduce a mapping label, which enables the single network to effectively learn specific representations between multiple source and target domains. Importantly, the trained network with various mapping labels can effectively correct not only amyloid PET datasets in multiple training domains but also unseen PET datasets obtained with new radiotracers, acquisition protocols, and PET scanners.

We performed quantitative and qualitative analyses to demonstrate that the proposed method for amyloid PET imaging can reduce scanning time while maintaining high diagnostic reliability.

\begin{figure*}[ht]
\centering
\includegraphics[width=1\textwidth]{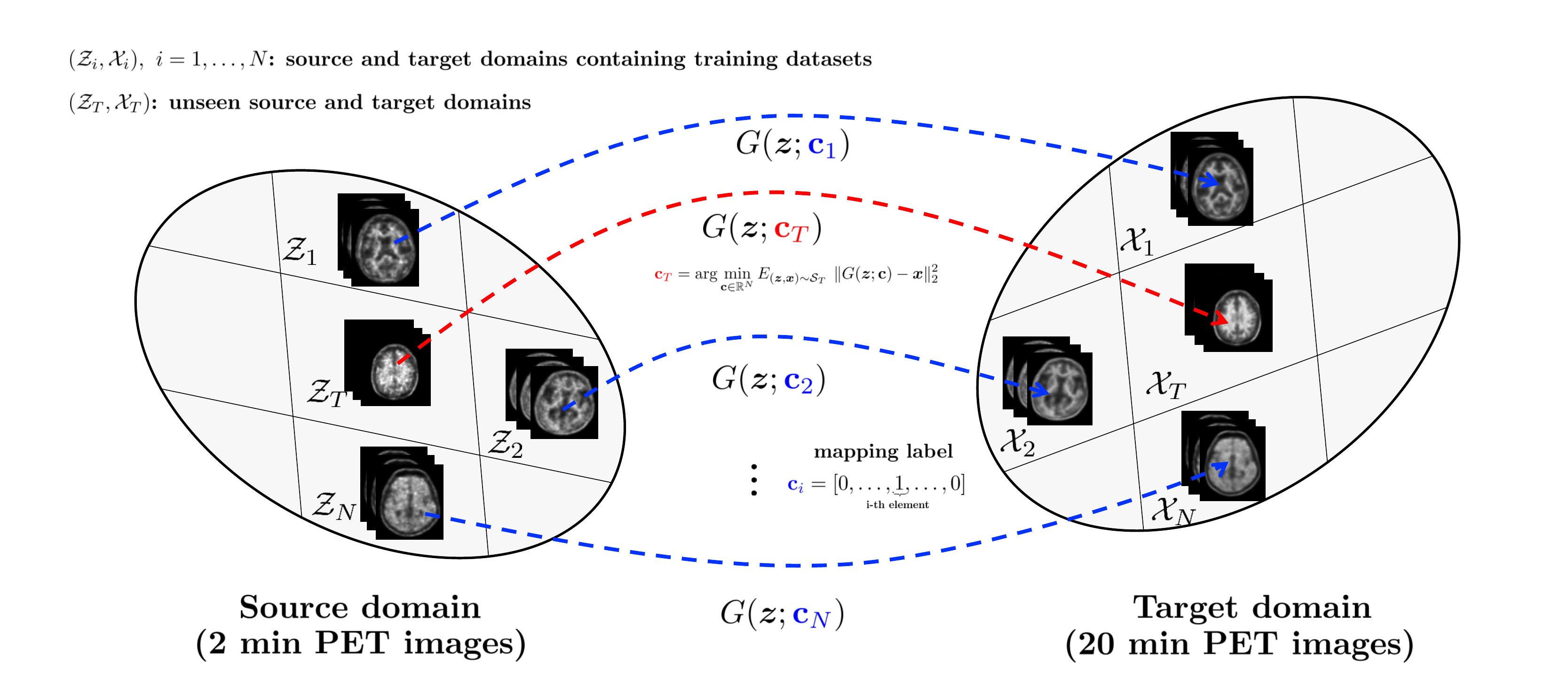}
\caption{Schematic diagram of proposed method for restoring short-scanning amyloid PET images. The mapping label $\c$ allows single network $G$ to effectively learn specific representations between $N$ different source and target domains $\mathcal{Z}_i$ and $\mathcal{X}_i, i=1,2,\ldots,N$. The network, trained with various mapping labels, has the capability to effectively correct amyloid PET datasets from an unseen domain $\mathcal{Z}_T$.}
\label{fig-main}
\end{figure*}

\section{Method}
Let $\z$ and $\x$ be amyloid PET images acquired from short and normal scan times, respectively. We assume that $\z =\x + \n$, where $\n$ is the noise artifact in $\z$ caused by the short-scanning time. The goal of this study is to learn a correction map $G: \z \to \x$ using $N$ different training datasets $\mS_i = \{(\z,\x)~:~\z\in\mathcal Z_i,~\x\in\mathcal X_i\},~i=1,2\ldots, N$. Here, $\mathcal Z_i$ and $\mathcal X_i$ are source and target domains that respectively exhibit different image characteristics depending on the acquisition environments, including the type of radiotracers used, the type of PET scanner, the type of reconstruction methods employed, and the scanning protocol.

Learning the map $G$ between two domains $\mathcal Z_i$ and $\mathcal X_i$ from the training datasets $\cup_{i=1}^N \mS_i$, which suffer from diverse perturbations, is challenging. To address this difficulty, we introduce mapping labels $\c_i,~i=1,2,\ldots, N$, which enable the $G:(\z;\c_i)\to \x$ maps $\z\in\mathcal Z_i$ to $\x\in\mathcal X_i$. We assigned the mapping label $\c_i$ as a one-hot vector of length $N$, where the $i$-th element was set to 1 and the other elements are set to 0.
Interestingly, the trained $G$ with these mapping labels possesses the capability to effectively map $\z$ to $\x$ sampled from the unseen source and target domains $\mathcal Z_T$ and $\mathcal X_T$, respectively. Specifically, we assumed that $\mathcal Z_T$ and $\mathcal X_T$ are sufficiently close to the source and target domains $\mathcal Z_i$ and $\mathcal X_i,~i=1,2,\ldots, N$, respectively. Then, for $\z\in\mathcal Z_T$, and $ \x\in\mathcal X_T$, there exists $\c_T\in\R^N$ such that $G(\z;\c_T) \approx \x$. Based on this observation, for the trained $G$ using $\cup_{i=1}^N \mS_i$, we estimated $\c_T$ by solving the following minimization problem:
\begin{align}\label{problem_c_t}
\arg\min_{\c\in \R^N} E_{(\z,\x)\sim\mS_T}~ \|G(\z;\c)-\x\|_2^2,
\end{align}
 where $E[\cdot]$ denotes the expectation and $\mS_T= \{(\z,\x)~:~\z\in\mathcal Z_T,~\x\in\mathcal X_T\}$. The schematic diagram of the proposed approach is illustrated in Fig. \ref{fig-main}.
\subsection{How to train network $G$ with the domain label $\c$}
 In this study, we incorporated the adaptive instance normalization (AdaIN) layer \cite{Huang2017} into the network $G$ to facilitate the conversion of feature maps from the $\mathcal Z_i$ to $\mathcal X_i$. We utilized the label $\mathbf{c}$ to generate the mean $\mu$ and variance $\sigma$ for the AdaIN layer, which are achieved through a mapping network $f$. Specifically, let $\h^{(k)}_j$ be a vectorized feature map of $G$ at the $j$-th channel of the $k$-th stage. Then, the AdaIN operation $\Gamma$ is applied as follows:
\begin{align}
\Gamma(\h^{(k)}_j, f^{(k)}(\c))=\f{\sigma_j^{(k)}}{\sigma(\h^{(k)}_j)} (\h^{(k)}_j-\mu(\h^{(k)}_j)\mathbf 1) + \mu_j^{(k)} \mathbf 1,
\end{align}
where $\sigma(\x)$ and $\mu(\x)$ respectively denote variance and mean of $\x$, and $\mathbf 1$ denotes a vector of ones whose every element is equal to one, and its size is same as that of $\h^{(k)}_j$. Here, $f^{(k)}$ is a mapping network at $k$-th stage that maps
\begin{align}
  f^{(k)}: \c \to \left[\begin{array}{cc}  \mu_1^{(k)} & \sigma_1^{(k)}\\ \vdots & \vdots \\ \mu_{J_k}^{(k)} & \sigma_{J_k}^{(k)}\\ \end{array}\right].
\end{align}
Here, $J_k$ is the number of channels of feature map at $k$-th stage.

In clinical practice, it is common to encounter data imbalance during training, where the number of collected samples in each domain is not equal. This can have a negative impact on the training performance of correction functions $G$. To address this issue, we used a weighted least squared loss that considers training samples $\cup_{i=1}^N \mS_i$ from $N$ different domains, as shown in eqn. \eref{wls}. However, using least-squares loss alone can lead to over-smoothed images with loss of details \cite{Jeong2021,Yang2018}. To overcome this limitation, we employed CycleGAN loss, which includes adversarial and cycle-consistency losses, in addition to the weighted least squared loss. The incorporation of CycleGAN loss in paired learning could significantly enhance the overall performance and quality of the generated images \cite{Xue2021}. In the proposed CycleGAN framework, the correction function $G$ is trained simultaneously with an additional backward function $F: (\x;\c)\to \z$, which maps the generated image back to the original domain $\mathcal Z$. Moreover, two discriminators $D_{\mathcal X}$ and $D_{\mathcal Z}$ are introduced to distinguish between the generated images and real images in their respective domains $\mathcal X$ and $\mathcal Z$.

In summary, the proposed correction function $G$ is trained along with the backward function $F$ and the discriminators $D_{\mathcal X}$ and $D_{\mathcal Z}$ by solving the following min-max problem:
\begin{align}
\argmin_{G,F}\argmax_{D_{\mathcal X}, D_{\mathcal Z}}~&\mL_{adv}(G,F, D_{\mathcal X}, D_{\mathcal Z}) + \lambda_1 \mL_{cyc}(G,F)\\ &+ \lambda_2 \mL_{wls}(G,F),
\end{align}
where
\begin{itemize}
\item $\mL_{adv}$ denotes the adversarial loss. To improve the training stability, we use the least-squares loss \cite{Mao2017}:
\begin{multline}
\mL_{adv}(G,F, D_{\mathcal X}, D_{\mathcal Z}) = \\
E_{(\x,\z)\sim\cup_{i=1}^N \mS_i}\left[ (D_{\mathcal X}(\x))^2 + (1-D_{\mathcal X}(G(\z;\c)))^2 \right.\\
\left. (D_{\mathcal Z}(\z))^2 + (1-D_{\mathcal Z}(F(\x;\c)))^2 \right].
\end{multline}
\item $\mL_{cyc}$ denotes the cycle-consistency loss given by
\begin{multline}
\mL_{cyc}(G,F) = \\
E_{(\x,\z)\sim\cup_{i=1}^N \mS_i}\left[\|F(G(\z;\c);\c)-\z\|_2^2 + \|G(F(\x;\c);\c)-\x\|_2^2\right].
\end{multline}
\item $\mL_{wls}$ denotes the weighted least squared loss given by
\begin{multline}\label{wls}
\mL_{wls}(G,F) = \\
E_{(\x,\z)\sim\cup_{i=1}^N \mS_i}\left[\|\w(G(\z;\c)-\x)\|_2^2 + \|\w(F(\x;\c)-\z)\|_2^2\right],
\end{multline}
where the weight is implemented as for each paired sample $(\z,\x)\in\mS_i$,
\begin{align}
\w_i = \f{\f{1}{|\mS_i |}}{\sum_{i=1}^N \f{1}{|\mS_i |}}.
\end{align}
Here, $|\mS|$ denotes the number of training samples in the set $\mS$.
\item $\lambda_1>0$ and $\lambda_2>0$ denote regularization parameters.
\end{itemize}

\subsection{Network architectures}

\begin{figure*}[ht]
\centering
\includegraphics[width=1\textwidth]{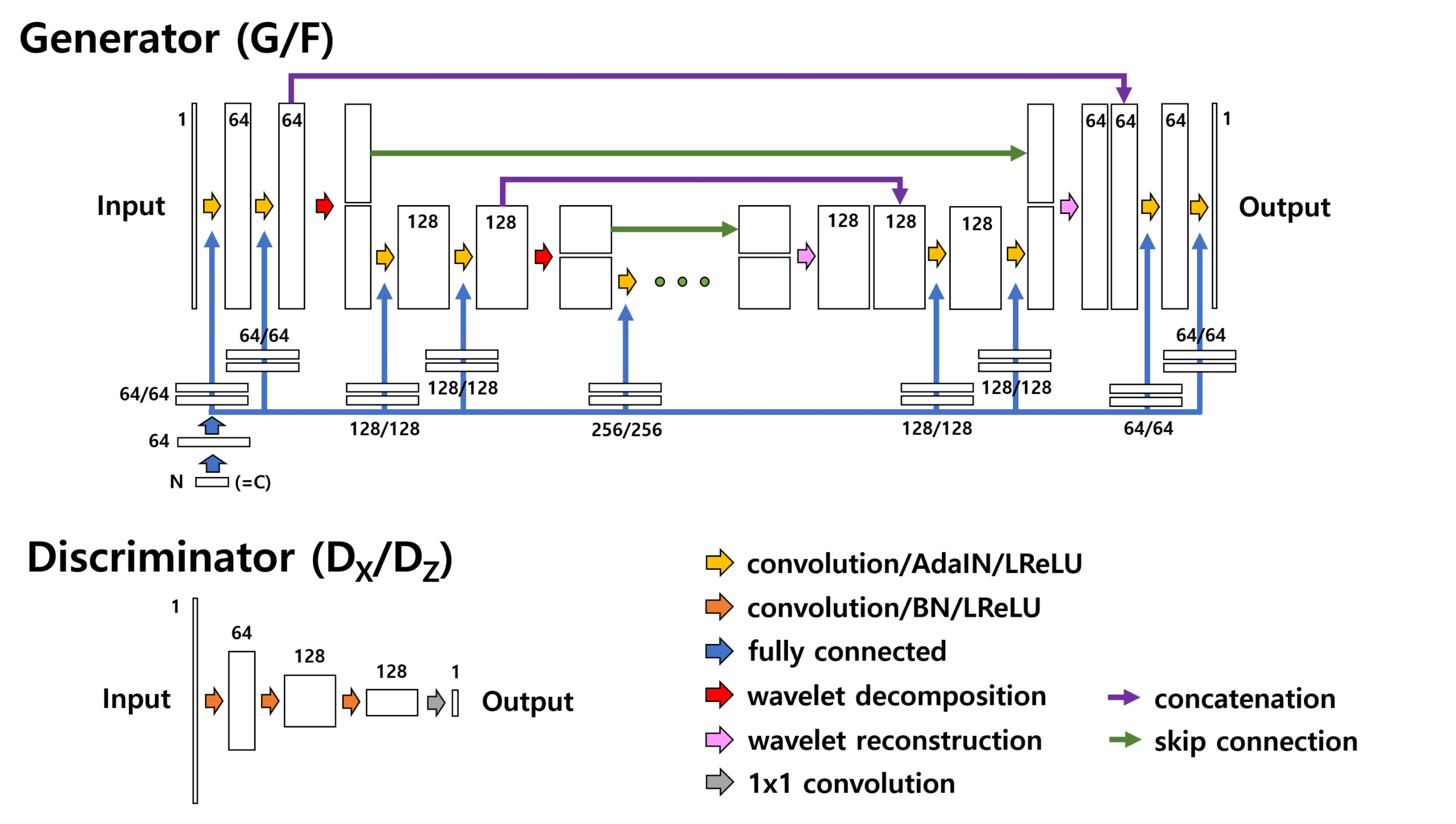}
\caption {Network architectures of the proposed method for restoring short-scanning amyloid PET images consist of generators and discriminators. Generators include two repeated $4\times4$ convolutions, each followed by AdaIN and leaky rectified linear unit. Mapping label $\c\in\R^N$ is passed through mapping networks to generate mean and variance for AdaIN layer. Discriminators were designed as a standard CNN without a fully connected layer. Numbers around the rectangular boxes indicate the number of channels.}
\label{fig_networks}
\end{figure*}

Network architectures of the proposed method are illustrated in Fig.~\ref{fig_networks}. The networks $G$ and $F$ were designed based on deep convolutional framelets \cite{Ye2018}, which were constructed with encoder-decoder structures and skipped connections. The networks consisted of two repeated $4\times4$ convolutions, each followed by the AdaIN and a leaky rectified linear unit (LReLU). In this setup, the mapping networks $f^{(k)}, k=1,\ldots,K$ that generate $J_k$ different $(\mu,\sigma)$ pairs were designed with two fully connected layers, each consisting of 64 nodes, and an output layer with $2J_k$ nodes. The generators employed 2-D Haar wavelet decomposition and reconstruction for down-sampling and up-sampling, respectively. In the encoder path, three high-pass filters after wavelet decomposition were directly skipped to the decoder path, while one low-pass filter was concatenated with the features in the decoder path at the same step. A convolution layer with a $1\times1$ window was added at the end to match the input and output image dimensions.

The discriminators $D_{\mathcal Z}$ and $D_{\mathcal X}$ were designed using a standard CNN without a fully connected layer. They consisted of three convolution layers, each with a $4\times4$ window and strides of two in each direction of the domain, followed by batch normalization~(BN) and a LReLU with a slope of 0.2. At the end of the architecture, a $1\times1$ convolution was added to generate a single-channel image.

\subsection{Implementation details}

We utilized the Adam optimization \cite{Kingma2014} to update the parameters of the networks for $G$, $F$, $D_{\mathcal X}$, and $D_{\mathcal Z}$. The learning rate for the optimization process was set to 0.0002. The initial weights of the networks were randomly initialized from a Gaussian distribution with a mean of 0 and a standard deviation of 0.01. We ran the training process for a total of 200 epochs. Training was implemented using TensorFlow~\cite{Abadi2016} with Horovod~\cite{horovod2018} on a 4-way GPU system (NVIDIA RTX 3090 24GB) with dual CPUs (Intel(R) Xeon Gold 6226R, 2.9GHz). With a mini-batch size of 32, the training took approximately 206 hours. In our study, the regularization parameters $\lambda_1$ and $\lambda_2$ were set to $10$ and $10$, respectively.

\begin{table*}[]
\caption{\label{tab_internal} Quantitative results of the correction methods for the internal validation. }
\centering
\begin{adjustbox}{width=\textwidth}

\begin{tabular}{|l|l|lll|lll|lll|}
\hline
\multirow{3}{*}{Training datasets}                                                        & \multirow{3}{*}{Methods}      & \multicolumn{3}{l|}{\multirow{2}{*}{FBB (n=119)}}                             & \multicolumn{3}{l|}{\multirow{2}{*}{FMM (n=30)}}                             & \multicolumn{3}{l|}{\multirow{2}{*}{FPN (n=15)}}                             \\
                                                                                          &                               & \multicolumn{3}{l|}{}                                                         & \multicolumn{3}{l|}{}                                                        & \multicolumn{3}{l|}{}                                                        \\ \cline{3-11}
                                                                                          &                               & \multicolumn{1}{l|}{NRMSE}              & \multicolumn{2}{l|}{SSIM}           & \multicolumn{1}{l|}{NRMSE}              & \multicolumn{2}{l|}{SSIM}           & \multicolumn{1}{l|}{NRMSE}              & \multicolumn{2}{l|}{SSIM}           \\ \hline
                                                                                          & 2 min image                   & \multicolumn{1}{l|}{15.613}				& \multicolumn{2}{l|}{0.839}          & \multicolumn{1}{l|}{20.494}                  & \multicolumn{2}{l|}{0.807}               & \multicolumn{1}{l|}{31.360}                  & \multicolumn{2}{l|}{0.798}               \\ \hline
\multirow{2}{*}{\begin{tabular}[c]{@{}l@{}}Single \\ ($\mS_i~,i=1,2,3$)\end{tabular}}     & {[}M1{]} adv+ls               & \multicolumn{1}{l|}{11.983}             & \multicolumn{2}{l|}{0.876}          & \multicolumn{1}{l|}{14.633}            & \multicolumn{2}{l|}{0.851}          & \multicolumn{1}{l|}{12.736}            & \multicolumn{2}{l|}{0.865}          \\ \cline{2-11}
                                                                                          & {[}M2{]} adv+ls+cycle         & \multicolumn{1}{l|}{10.921}             & \multicolumn{2}{l|}{0.893}          & \multicolumn{1}{l|}{13.736}            & \multicolumn{2}{l|}{\textbf{0.869}} & \multicolumn{1}{l|}{12.607}            & \multicolumn{2}{l|}{0.877}          \\ \hline
\multirow{3}{*}{\begin{tabular}[c]{@{}l@{}}Multiple\\ ($\cup_{i=1}^3\mS_i$)\end{tabular}} & {[}M3{]} adv+wls              & \multicolumn{1}{l|}{12.846}             & \multicolumn{2}{l|}{0.885}          & \multicolumn{1}{l|}{14.038}            & \multicolumn{2}{l|}{0.859}          & \multicolumn{1}{l|}{13.226}            & \multicolumn{2}{l|}{0.858}          \\ \cline{2-11}
                                                                                          & {[}M4{]} adv+wls+cycle        & \multicolumn{1}{l|}{11.956}             & \multicolumn{2}{l|}{0.879}          & \multicolumn{1}{l|}{\textbf{13.558}}   & \multicolumn{2}{l|}{0.855}          & \multicolumn{1}{l|}{14.386}            & \multicolumn{2}{l|}{0.854}          \\ \cline{2-11}
                                                                                          & {[}ours{]} adv+wls+cycle+$\c$ & \multicolumn{1}{l|}{\textbf{10.837}}    & \multicolumn{2}{l|}{\textbf{0.894}} & \multicolumn{1}{l|}{13.589}            & \multicolumn{2}{l|}{0.867}          & \multicolumn{1}{l|}{\textbf{12.131}}   & \multicolumn{2}{l|}{\textbf{0.880}} \\ \hline
\end{tabular}
\end{adjustbox}

\end{table*}

\begin{table*}[]
\caption{\label{tab_temporal} Quantitative results of the correction methods for the temporal validation. }
\centering
\begin{adjustbox}{width=\textwidth}

\begin{tabular}{|l|l|lll|lll|lll|}
\hline
\multirow{3}{*}{Training datasets}                                                     & \multirow{3}{*}{Methods}               & \multicolumn{3}{l|}{\multirow{2}{*}{FBB (n=20)}}                                           & \multicolumn{3}{l|}{\multirow{2}{*}{FMM (n=20)}}                                                    & \multicolumn{3}{l|}{\multirow{2}{*}{FPN (n=20)}}                                           \\
                                                                                       &                                        & \multicolumn{3}{l|}{}                                                                      & \multicolumn{3}{l|}{}                                                                               & \multicolumn{3}{l|}{}                                                                      \\ \cline{3-11}
                                                                                       &                                        & \multicolumn{1}{l|}{NRMSE}                    & \multicolumn{2}{l|}{SSIM}                   & \multicolumn{1}{l|}{NRMSE}                    & \multicolumn{2}{l|}{SSIM}                           & \multicolumn{1}{l|}{NRMSE}                    & \multicolumn{2}{l|}{SSIM}                   \\ \hline
                                                                                       & 2 min image                            & \multicolumn{1}{l|}{15.179}                        & \multicolumn{2}{l|}{0.823}                       & \multicolumn{1}{l|}{19.882}                        & \multicolumn{2}{l|}{0.811}                          & \multicolumn{1}{l|}{34.934}                        & \multicolumn{2}{l|}{0.815}                       \\ \hline
\multirow{2}{*}{\begin{tabular}[c]{@{}l@{}}Single \\ ($\mS_i~,i=1,2,3$)\end{tabular}}  & {[}M1{]} adv+ls                         & \multicolumn{1}{l|}{11.626}                  & \multicolumn{2}{l|}{0.866}                  & \multicolumn{1}{l|}{13.615}                  & \multicolumn{2}{l|}{0.859}                           & \multicolumn{1}{l|}{12.367}                  & \multicolumn{2}{l|}{0.865}                  \\ \cline{2-11}
                                                                                       & {[}M2{]} adv+ls+cycle                   & \multicolumn{1}{l|}{10.648}                  & \multicolumn{2}{l|}{0.886}                  & \multicolumn{1}{l|}{13.058}                  & \multicolumn{2}{l|}{0.873}                           & \multicolumn{1}{l|}{12.493}                  & \multicolumn{2}{l|}{0.882}                  \\ \hline
\multirow{3}{*}{\begin{tabular}[c]{@{}l@{}}Multiple\\ ($\cup_{i=1}^3\mS_i$)\end{tabular}} & \multirow{2}{*}{{[}M3{]} adv+wls+cycle} & \multicolumn{1}{l|}{\multirow{2}{*}{11.984}} & \multicolumn{2}{l|}{\multirow{2}{*}{0.886}} & \multicolumn{1}{l|}{\multirow{2}{*}{12.844}} & \multicolumn{2}{l|}{\multirow{2}{*}{\textbf{0.876}}} & \multicolumn{1}{l|}{\multirow{2}{*}{13.866}} & \multicolumn{2}{l|}{\multirow{2}{*}{0.881}} \\
                                                                                       &                                        & \multicolumn{1}{l|}{}                        & \multicolumn{2}{l|}{}                       & \multicolumn{1}{l|}{}                        & \multicolumn{2}{l|}{}                                & \multicolumn{1}{l|}{}                        & \multicolumn{2}{l|}{}                       \\ \cline{2-11}
                                                                                       & {[}ours{]} adv+wls+cycle+$\c$           & \multicolumn{1}{l|}{\textbf{10.557}}         & \multicolumn{2}{l|}{\textbf{0.888}}         & \multicolumn{1}{l|}{\textbf{12.820}}         & \multicolumn{2}{l|}{0.873}                           & \multicolumn{1}{l|}{\textbf{11.763}}         & \multicolumn{2}{l|}{\textbf{0.886}}         \\ \hline
\end{tabular}
\end{adjustbox}

\end{table*}

\section{Experiments}
The retrospective study protocol (DAUHIRB-22-053) was reviewed and approved by the Institutional Review Board (IRB) of Dong-A University Hospital (DAUH). Informed consentsince only anonymized data was used for research purposes. All methods used in this study were in accordance with relevant guidelines and regulations. Some of the data used in this article were obtained from the Alzheimer's Disease Neuroimaging Initiative (ADNI) database (adni.loni.usc.edu). The ADNI was launched in 2003 as a public-private partnership, led by Principal Investigator Michael W. Weiner, MD. The primary goal of ADNI has been to test whether serial magnetic resonance imaging (MRI), positron emission tomography (PET), other biological markers, and clinical and neuropsychological assessment can be combined to measure the progression of mild cognitive impairment and early Alzheimer's disease. 
For up-to-date information, see \url{www.adni-info.org}.

\subsection{Datasets}
In this study, a total of 814 FBB-PET, 415 FMM-PET, and 208 FPN-PET scans were collected from the three different domains. These scans were divided into training, internal validation, and temporal validation sets. Specifically, 734, 60, and 20 scans were used for training, internal validation, and temporal validation for FBB-PET, respectively. Similarly, 365, 30, and 20 scans were used for FMM-PET, and 173, 15, and 20 scans were used for FPN-PET. Each scan consisted of 110 images of size $400\times 400$. The training and internal validation datasets were obtained between December 2015 and December 2021, while the temporal validation dataset was obtained between May and September 2022. All FBB-, FMM-, and FPN-PET scans were acquired using a Biograph mCT flow scanner (Siemens Healthcare, Knoxville, TN, USA). The PET/CT imaging was conducted in accordance with the standard protocol of DAUH, which is consistent with the method used in previous studies \cite{Jeong2017,Jeong2021}. In each scan protocol, the patients were intravenously injected with 300 MBq of FBB, 185 MBq of FMM, and 370 MBq of FPN. PET images were acquired 90 minutes after injection for FBB- and FMM-PET, and 60 minutes for FPN-PET. The PET scans covered the skull vertex to the skull base and were acquired for 20 minutes using 3-dimensional and list modes. The ground-truth $\x$ and short-scanning static PET image $\z$ were reconstructed from the 20-minute and first 2-minute scan data, respectively, with the same parameters. The paired training samples for FBB-, FMM-, and FPN-PET were denoted as $\mS_1$, $\mS_2$, and $\mS_3$, respectively, and the mapping labels were assigned as $\c_1 = [1,~0,~0]$ for FBB-PET, $\c_2 = [0,~1,~0]$ for FMM-PET, and $\c_3=[0,~0,~1]$ for FPN-PET.

For external validations, we collected 54 FBB-PET and 30 FBP-PET scans from the ADNI database. These scans contained a series of $4 \times 5$ min PET images. Note that the proposed model, trained on the dataset of paired 2-minute and 20-minute images, was tested on the first 5-minute PET images. Here, a Gaussian filter with 4~mm FWHM was applied to all FBB-PET images of the ADNI datasets. Among the collected ADNI scans, 8 FBB-PET and 6 FPN-PET scans were randomly selected for estimation of unknown label $\c_T$, while the remaining scans were employed for external validations.

\begin{figure*}[ht]
\centering
\includegraphics[width=1.0\textwidth]{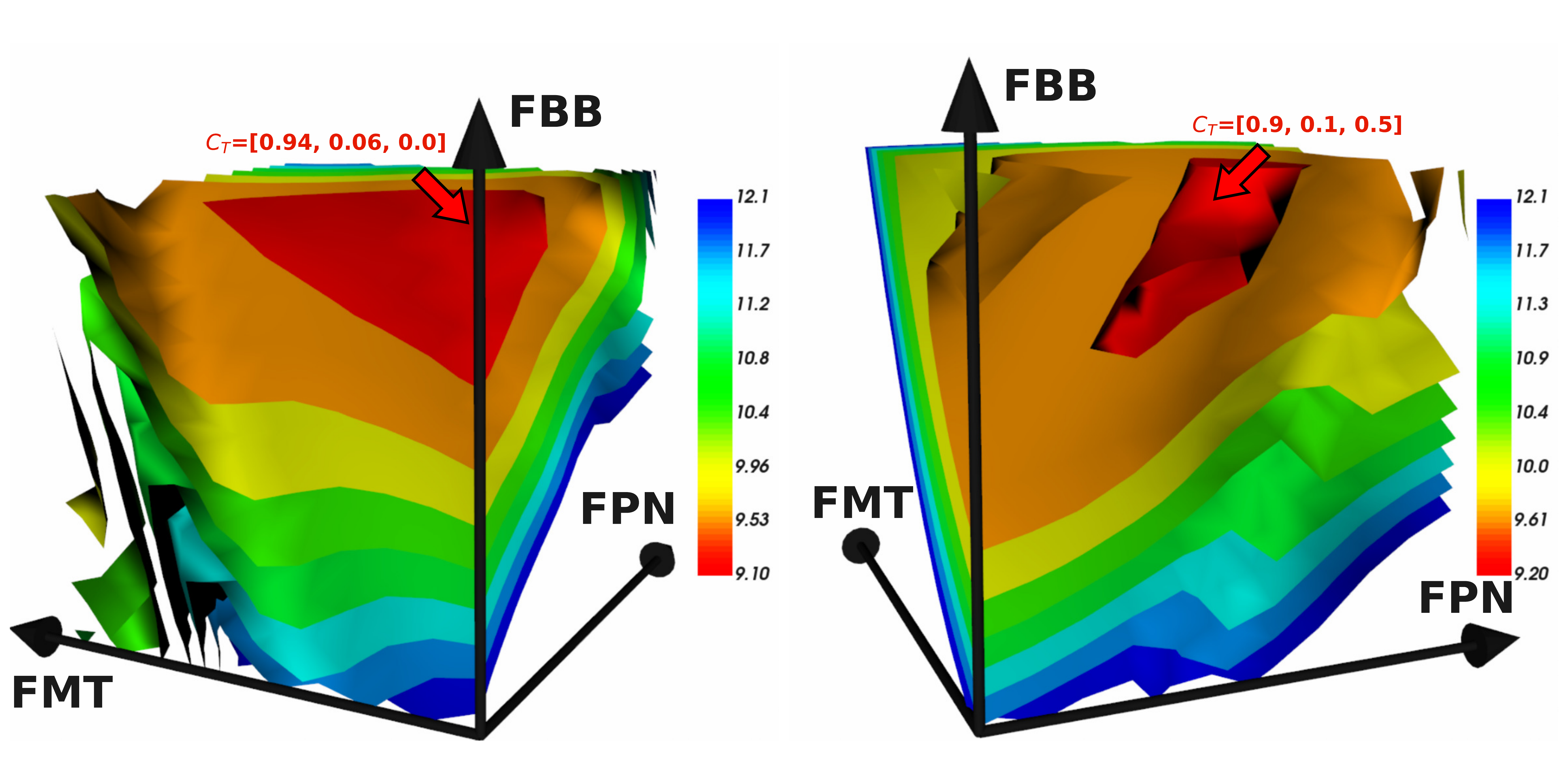}
\caption{Illustration of mapping labels $\c_T$ for ADNI-FBB (left figure) and ADNI-FBP datasets (right figure). Color indicates value of the objective function in \eref{problem_c_t}.}
\label{fig-mappinglabel}
\end{figure*}

In our study, we determined the value of $\c_T$ as follows: first, we restricted the domain $\R^3$ to $[-\eps,1+\eps]^3$, where $\eps$ is a small positive value. Next, we discretized the restricted domain with a linear spacing and computed the objective function defined in equation \eref{problem_c_t} for each point within the discrete domain. Finally, we selected the point with the minimum objective function value as our estimate for $\c_T$. The computed objective function values on the domain $[-\eps,1+\eps]^3$ are visually presented in Fig. \ref{fig-mappinglabel}.

\begin{table}[]
\caption{\label{tab_external} Quantitative results of the correction methods for external validation.}
\centering
\begin{adjustbox}{width=\textwidth}

\begin{tabular}{|l|l|lll|lll|ll}
\cline{1-8}
\multirow{2}{*}{Training datasets}                                                       & \multirow{2}{*}{Methods}    																								& \multicolumn{3}{l|}{ADNI-FBB (n=46)}                      & \multicolumn{3}{l|}{ADNI-FBP (n=24)}                      &  &  \\ \cline{3-8}
                                                                                         &                                                                                                                           & \multicolumn{1}{l|}{NRMSE}    & \multicolumn{2}{l|}{SSIM}  & \multicolumn{1}{l|}{NRMSE}    & \multicolumn{2}{l|}{SSIM}  &  &  \\ \cline{1-8}
                                                                                         & 5 min images                                                                                                              & \multicolumn{1}{l|}{11.963}  & \multicolumn{2}{l|}{0.861} & \multicolumn{1}{l|}{16.117}  & \multicolumn{2}{l|}{0.808} &  &  \\ \cline{1-8}
FBB $(\mS_1)$                                                                            & {[}M1{]} adv+ls                                                                                                            & \multicolumn{1}{l|}{10.775}  & \multicolumn{2}{l|}{0.867} & \multicolumn{1}{l|}{11.386}       & \multicolumn{2}{l|}{0.861}     &  &  \\ \cline{1-8}
\multirow{2}{*}{\begin{tabular}[c]{@{}l@{}}Multiple\\ $(\sum_{i=1}^3 \mS_i)$\end{tabular}} & {[}M2{]} adv+wls+cycle                                                                                                     & \multicolumn{1}{l|}{9.597}   & \multicolumn{2}{l|}{0.893} & \multicolumn{1}{l|}{11.974}        & \multicolumn{2}{l|}{0.891}      &  &  \\ \cline{2-8}
                                                                                         & \begin{tabular}[c]{@{}l@{}}[ours] adv+wls+cycle+$\c_T$\\ \scriptsize {$\c_T =[0.94,0.06,0]$ for FBB}\\ \scriptsize{$\c_T=[0.9,0.1,0.5]$ for FBP}\end{tabular} & \multicolumn{1}{l|}{\bf 8.218}   & \multicolumn{2}{l|}{\bf 0.899} & \multicolumn{1}{l|}{\bf 8.251}   & \multicolumn{2}{l|}{ 0.897} &  &  \\ \cline{1-8}
\multirow{2}{*}{Training datasets}                                                       & \multirow{2}{*}{Methods}                                                                                                  & \multicolumn{3}{l|}{ADNI-FBB (n=46)}                      & \multicolumn{3}{l|}{ADNI-FBP (n=24)}                      &  &  \\ \cline{3-8}
                                                                                         &                                                                                                                           & \multicolumn{1}{l|}{NRMSE}    & \multicolumn{2}{l|}{SSIM}  & \multicolumn{1}{l|}{NRMSE}    & \multicolumn{2}{l|}{SSIM}  &  &  \\ \cline{1-8}
ADNI-FBB (n=54)                                                                          & {[}M3{]} adv+wls+cycle                                                                                                     & \multicolumn{1}{l|}{8.490}        & \multicolumn{2}{l|}{0.897}      & \multicolumn{1}{c|}{-}       & \multicolumn{2}{c|}{-}     &  &  \\ \cline{1-8}
ADNI-FBP (n=30)                                                                          & {[}M4{]} adv+wls+cycle                                                                                                     & \multicolumn{1}{c|}{-}       & \multicolumn{2}{c|}{-}     & \multicolumn{1}{l|}{8.910}   & \multicolumn{2}{l|}{\bf 0.899} &  &  \\ \cline{1-8}
\end{tabular}
\end{adjustbox}

\end{table}

\subsection{Quantitative, statistical, and qualitative analyses}
Image quality was evaluated using normalized root mean-square error (NRMSE) and structure similarity index measure (SSIM) metrics. Precisely, for two vectors $\z$ and $\x$, the NRMSE is computed as follows:
\begin{align}
  NRMSE(\z,\x) = \sqrt{\f{\|\z-\x\|_2^2}{\|\x\|_2^2}}\times 100,
\end{align}
where $\|\cdot\|_2$ denotes the standard $l_2$ norm.

The SSIM is computed as follows:
\begin{align}
  SSIM(\z,\x) = \f{(2\mu_{\z}\mu_{\x}+\alpha_1)(2\sigma_{\z\x}+\alpha_2)}{(\mu_{\z}^2+\mu_{\x}^2+\alpha_1)(\sigma_{\x}^2+\sigma_{\x}^2+\alpha_2)},
\end{align}
where $\alpha_1$ and $\alpha_2$ are parameters. We chose $\alpha_1=(0.0002 \times 65535)^2$ and $\alpha_2=(0.0007 \times 65535)^2$ based on the experimental results reported in our previous study \cite{Jeong2021}.

We also computed the standardized uptake value ratio (SUVR) using PMOD 3.6 software (PMOD Technologies, Zurich, Switzerland) \cite{Bullich2017}. PET images were normalized spatially using the transformation matrix of each participant and were applied to an automated anatomical labeling template of PMOD. We normalized spatially and applied all pairs of corrected and ground-truth PET images (i.e., 20-min PET images) to the amyloid cortical composite atlas in the PMOD program. The cortical composite was set up as the frontal, temporal, and parietal lobes and cingulate of the brain, and we reconstructed the volume-of-interests of the atlas to calculate the SUVRs of the representative areas. The reference region used for SUVR calculation was the whole cerebellum. Finally, we compared the SUVRs of the identical areas between the corrected and ground-truth PET images of the external validation dataset (ADNI-FBB and ADNI-FBP).

In addition, we performed visual interpretation to evaluate the clinical applicability of the corrected PET images. Following the conventional interpretation protocol, two nuclear medicine physicians certified and experienced in amyloid PET reading independently determined the amyloid status (positive or negative) for both corrected and ground-truth PET images. They were blinded to the clinical data and read all PET images of the external validation dataset. We measured the accuracy of the visual interpretation for corrected PET images for each physician and analyzed the intra-observer agreements of binary classification of corrected and ground-truth PET images for each physician.

\begin{figure*}[ht]
\centering
\includegraphics[width=1.0\textwidth]{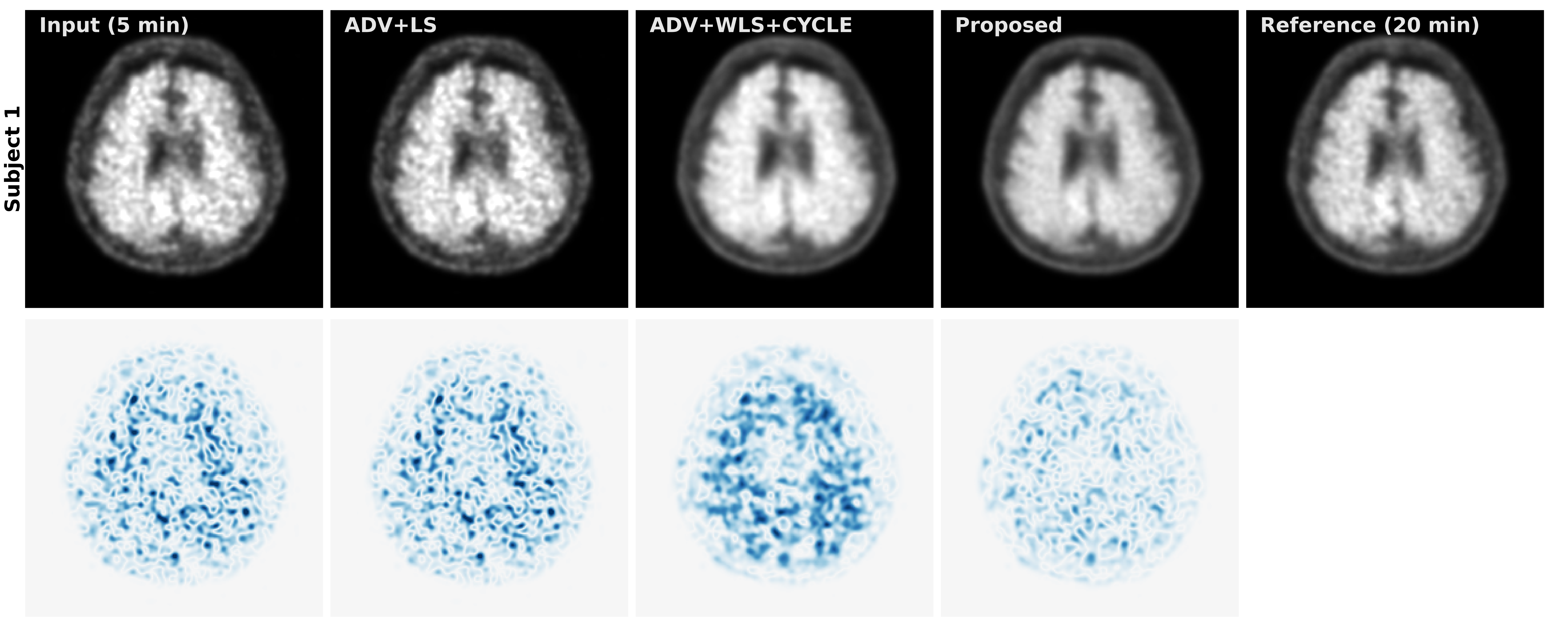}
\includegraphics[width=1.0\textwidth]{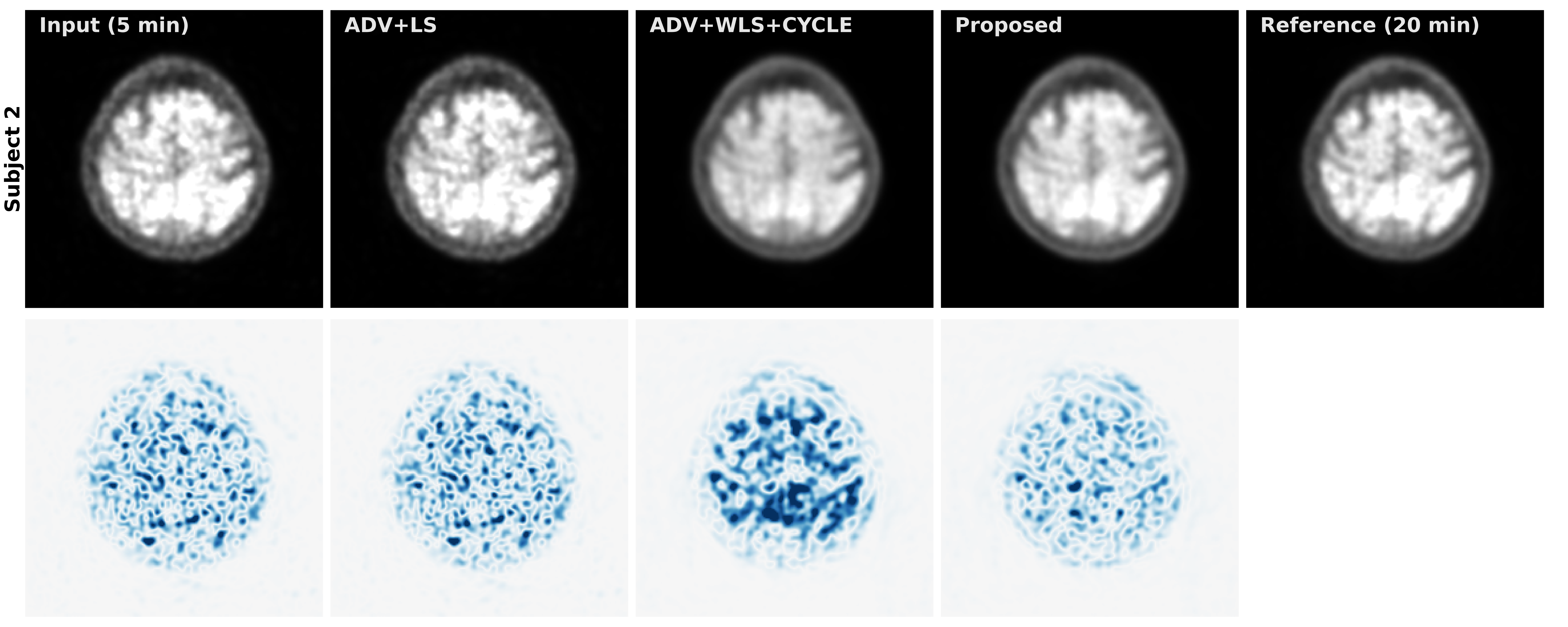}
\caption{Correction results for external validation (ADNI-FBB). Each column (from left to right) shows the input (5 min), adv+ls [M1], adv+wls+cycle [M2], proposed and reference (20 min) images. The second and fourth columns show difference images from the reference (20 min) image. WW/WC = [7000, 3500] for PET images, and WW/WC = [1200, 600] for difference images.}
\label{fig-external-fbb}
\end{figure*}

\subsection{Correction results}
\subsubsection{Internal validations}
Table \ref{tab_internal} illustrates the quantitative results of correction methods for internal validation datasets. For quantitative evaluation, we calculated NRMSE and SSIM between corrected images and the corresponding ground-truth images. In Table \ref{tab_internal}, the `M1' to `M4' denote the methods trained using the different combination of the losses such as  $\mL_{adv},~\mL_{ls},~\mL_{wls},~\mL_{cycle}$, and the different training datasets such as single and multiple training datasets. We note that Method 1 is the method used in our pervious study \cite{Jeong2021}.

The results demonstrate that Method 2 outperformed Method 1 for the FBB, FMM, and FPN datasets, respectively, suggesting that the cycle-consistency loss led to improved performance for each individual training dataset. However, Methods 3 and 4 performed worse than Methods 1 and 2, likely due to the added complexity of dealing with multiple source and target domains $\mathcal Z_i$ and $\mathcal X_i$, where $i=1,2,3$. The proposed method using mapping labels achieved the best performance for the FBB and FPN datasets and was close to the best performance for the FMM dataset. These findings suggest that the proposed network using mapping labels stably learned the relationships between multiple source and target domains. Furthermore, the proposed method achieved comparable or better performance compared to methods trained using only a single source and target domain (Method 2).

Table \ref{tab_temporal} summarizes the quantitative results of the correction methods for temporal validation dataset. The results were consistent with the those obtained from the internal validation dataset. Overall, the proposed method achieved the best performance, except for the SSIM value of the FMM dataset.

\begin{figure*}[ht]
\centering
\includegraphics[width=1.0\textwidth]{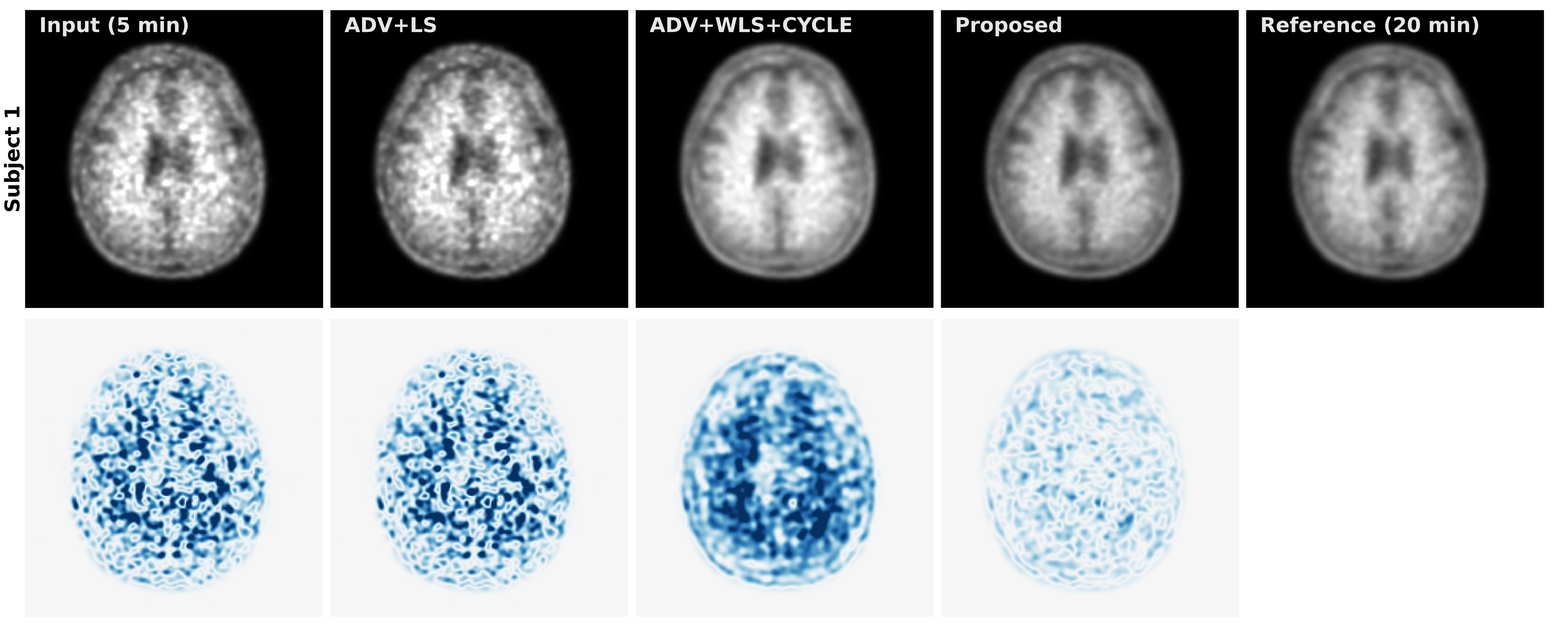}
\includegraphics[width=1.0\textwidth]{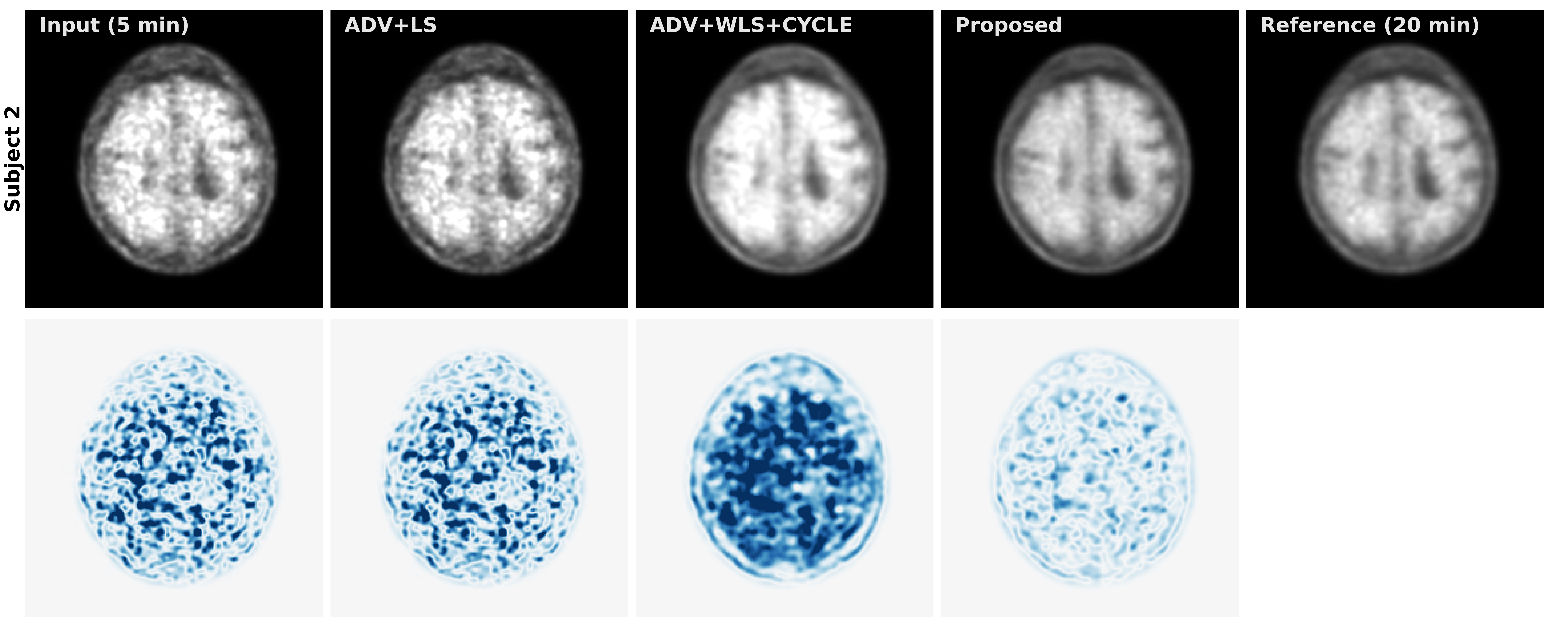}
\caption{Correction results for external validation (ADNI-FBP). Each column (from left to right) shows the input (5 min), adv+ls [M1], adv+wls+cycle [M2], proposed and reference (20 min) images. The second and fourth columns show difference images from the reference (20 min) image. WW/WC = [7000, 3500] for PET images, and WW/WC = [1200, 600] for difference images.}
\label{fig-external-fbp}
\end{figure*}

\subsubsection{External validations}
We performed an external validation of our proposed method by applying it to the ADNI-FBB and ADNI-FBP datasets, which were obtained from unseen domains. The estimated $\c_T$ for ADNI-FBB was found to be $[0.94, 0.06, 0]$, which is very close to the $\c_1=[1,0,0]$ assigned for the FBB dataset. On the other hand, the estimated $\c_T$ for ADNI-FBP was $[0.9,0.1,0.5]$. Since the FBP dataset was not used in training, it was represented as a suitable linear combination of $\c_1$, $\c_2$, and $\c_3$. We presented the quantitative results in Table \ref{tab_external}. Method 1 was trained using the FBB dataset ($\mS_1$), while Method 2 and the proposed method were trained using all FBB ($\mS_1$), FMM ($\mS_2$), and FPN ($\mS_3$) datasets. The proposed method, with the mapping label $\c_T$, outperformed Methods 1 and 2 for both ADNI-FBB and ADNI-FBP datasets. To further evaluate our proposed method, we compared its performance with Methods 3 and 4, which were trained using the 54 ADNI-FBB and 30 ADNI-FBP datasets used for external validation, respectively. Due to the limited number of datasets, we calculated the NRMSE and SSIM on the training datasets, and the results are presented in Table \ref{tab_external}. Even when compared with the training results of Methods 3 and 4, the proposed method achieved better or comparable results. We showed the visual comparisons in Figs.~\ref{fig-external-fbb} and \ref{fig-external-fbp}.

\begin{figure*}[ht]
\centering
\includegraphics[width=0.6\textwidth]{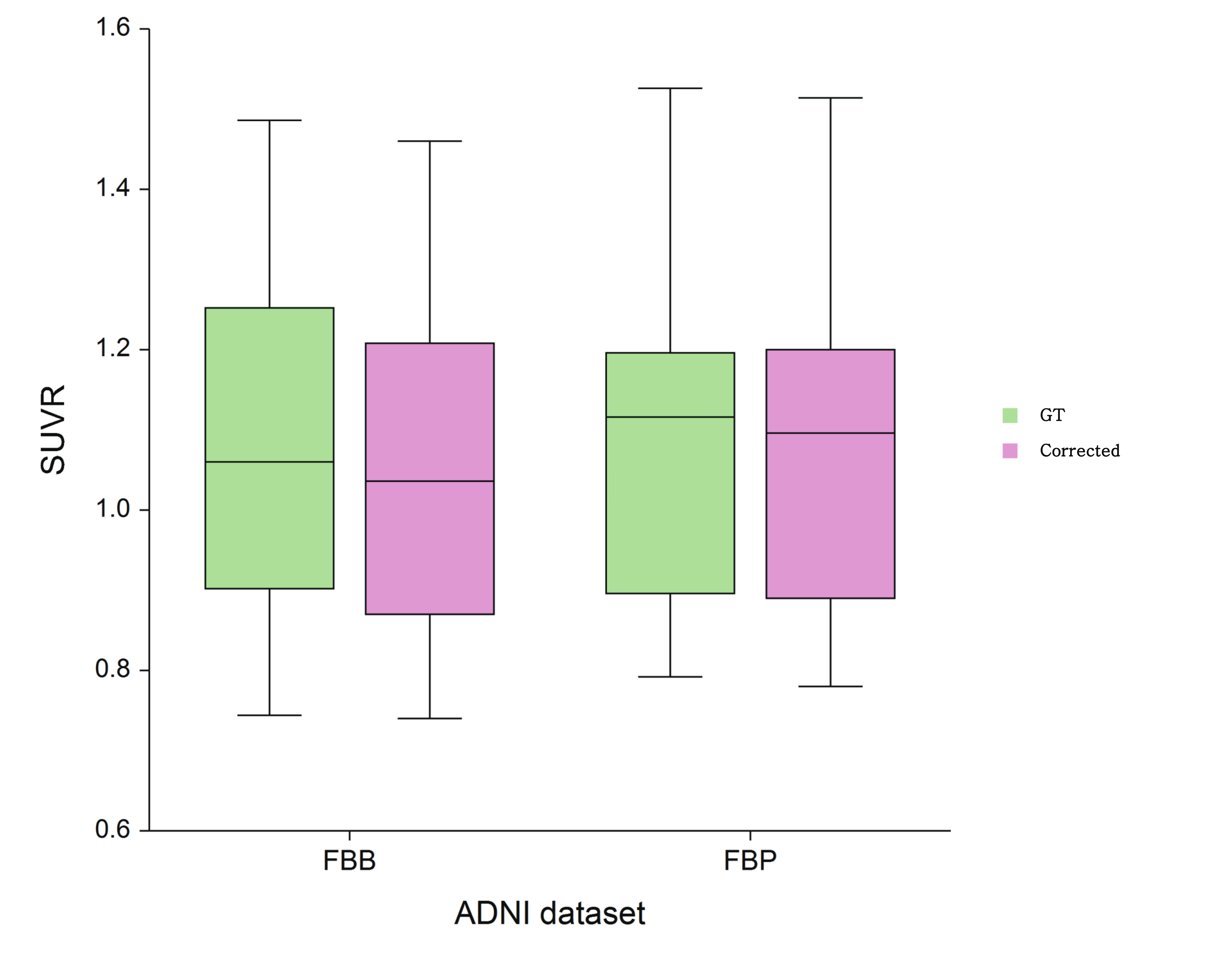}
\caption{Comparison of mean SUVRs in ground-truth (GT) and corrected PET images of external validation.}
\label{fig-suvr}
\end{figure*}

\begin{figure*}[ht]
\centering
\includegraphics[width=1.0\textwidth]{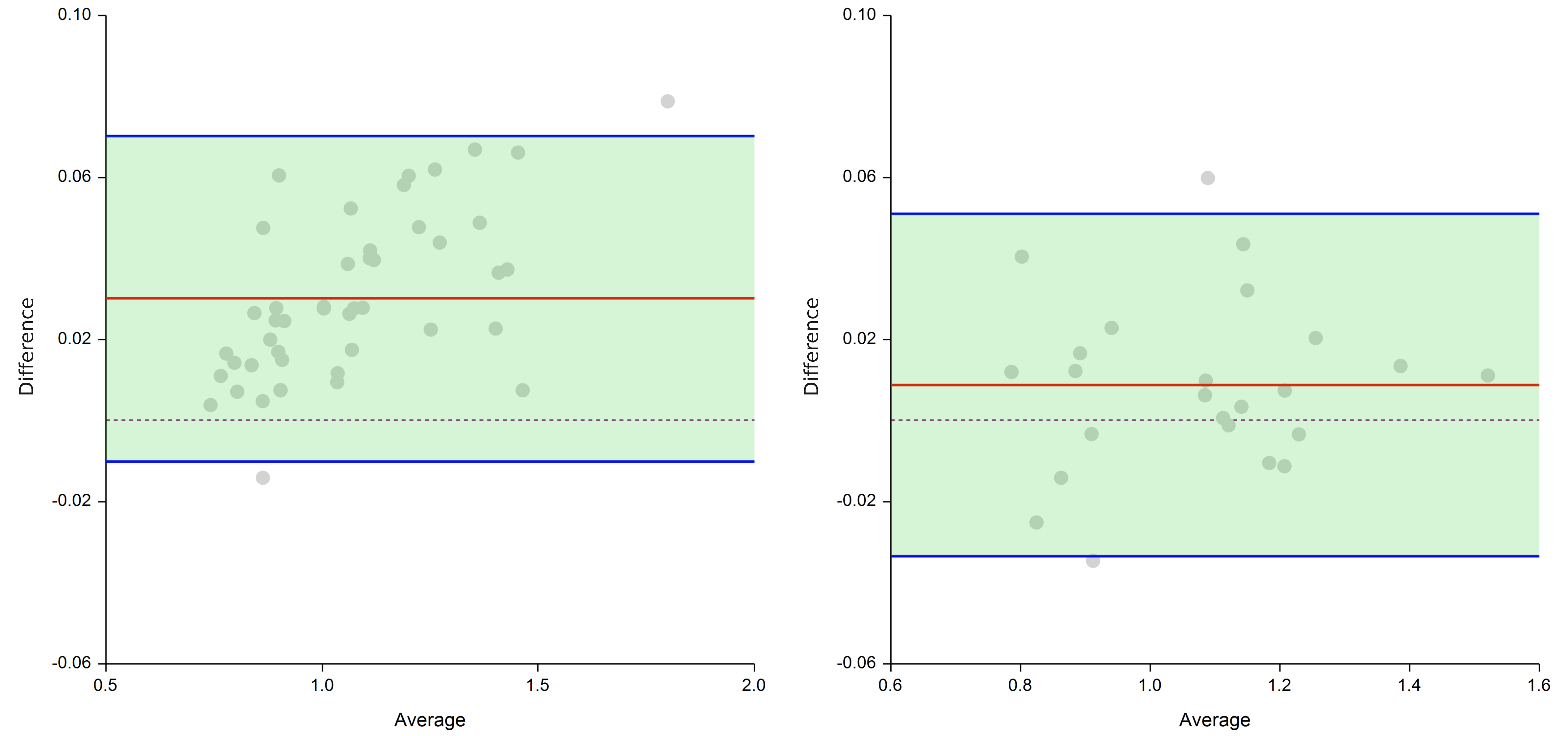}
\caption{The Bland-Altman graphs for ADNI-FBB (left figure) and ADNI-FBP (right figure) show the differences between ground-truth and corrected PET images. Red line represents average difference between the ground-truth and corrected PET images. Blue lines represent the upper and lower 95 \% confidence interval limits for the average difference.}
\label{fig-baplots}
\end{figure*}

We compared the range of SUVRs in the identical cortical composite area between the corrected image and the corresponding ground-truth image for both ADNI-FBB and ADNI-FBP datasets. As shown in Fig.~\ref{fig-suvr}, no statistically significant difference was revealed in the SUVRs of the cortical composite areas between ground-truth and corrected images for both cases. There was a very strong positive correlation in the ADNI-FBB dataset ($r = 0.998$, $p < 0.001$) and the ADNI-FBP dataset ($r = 0.994$, $p < 0.001$). We also evaluated the difference in the SUVRs between the two PET images. In the Bland-Altman analysis, the mean difference of SUVR between ground-truth and corrected images was 0.030 (95\% confidence interval (CI) -0.024, 0.036) in the ADNI-FBB dataset and 0.009 (95\% CI -0.000, 0.018) in the ADNI-FBP dataset. Upper and lower limits of agreement were 0.070 (95\% CI 0.060, 0.081) and -0.010 (95\% CI -0.021, 0.000) in the ADNI-FBB dataset, and 0.051 (95\% CI 0.035, 0.067) and -0.033 (95\% CI -0.049, -0.018) in the ADNI-FBP dataset, respectively. The Bland-Altman graphs for ADNI-FBB and ADNI-FBP were shown in Fig.~\ref{fig-baplots}.

The accuracy of visual binary classification for corrected PET images of the external validation set was 0.960 (reader 1), 0.920 (reader 2) for ADNI-FBP-PET images, and 0.980 (reader 1), 0.940 (reader 2) for ADNI-FBB-PET images. Additionally, we evaluated the intra-observer agreement using Cohen's weighted kappa by comparing visual interpretations between ground-truth and corrected PET images. The computed Cohen's weighted kappa was 0.911 (reader 1) and 0.826 (reader 2) for ADNI-FBP-PET images, and 0.959 (reader 1) and 0.874 (reader 2) for ADNI-FBB-PET images.

\section{Discussion and conclusion}

In this paper, we proposed a deep learning-based method for short-scanning amyloid PET image restoration that can be generalized across various image domains. The proposed method learns a correction map using amyloid PET images sampled from multiple source and target domains. Our main contribution is the introduction of a mapping label that facilitates the effective learning of specific representations between these domains. Furthermore, the mapping label enables the effective correction of not only amyloid PET datasets in multiple training domains but also PET datasets from unseen domains.

Our experiments demonstrate that the proposed method has the potential to reduce scanning time while maintaining high diagnostic reliability. The proposed method effectively restores short scan amyloid PET images from multiple training source domains, as evidenced by quantitative metrics such as NRMSE and SSIM. For the external validation datasets (ADNI-FBB and ADNI-FBP), the proposed method achieved comparable or even superior results compared to methods trained using the datasets used for external validations. Additionally, there was no statistically significant difference in the SUVRs of the cortical composite areas between the ground-truth and corrected images $(p < 0.001)$. The accuracy for determining amyloid status on the corrected PET images was high, with values of 0.970 for reader 1 and 0.930 for reader 2.

In our study, although the proposed method was trained using FBB, FMM, and FPN datasets from three different domains $(N=3)$, it can be easily generalized with the mapping labels $\c_i,~i=1,2,\ldots,N$ for large $N$. For external validation, various optimization methods can be applied to find the optimal $\c_T$ in the problem of equation \eref{problem_c_t}. The proposed method, which utilizes domain labels for short-scanning amyloid PET image restoration, can be a promising approach to overcoming the challenges of multi-domain learning in PET imaging.

\section*{Acknowledgments}
This work was supported by the National Institute for Mathematical Sciences(NIMS) grant funded by the Korean government (No. NIMS-B23910000). Y.J.J was supported by the National Research Foundation of Korea(NRF) grant funded by the Korea government(MSIT) (No. 2022R1A2C1005016). Data collection and sharing for this project was funded by the Alzheimer's Disease Neuroimaging Initiative (ADNI) (National Institutes of Health Grant U01 AG024904) and DOD ADNI (Department of Defense award number W81XWH-12-2-0012). ADNI is funded by the National Institute on Aging, the National Institute of Biomedical Imaging and Bioengineering, and through generous contributions from the following: AbbVie, Alzheimer's Association; Alzheimer's Drug Discovery Foundation; Araclon Biotech; BioClinica, Inc.; Biogen; Bristol-Myers Squibb Company; CereSpir, Inc.; Cogstate; Eisai Inc.; Elan Pharmaceuticals, Inc.; Eli Lilly and Company; EuroImmun; F. Hoffmann-La Roche Ltd and its affiliated company Genentech, Inc.; Fujirebio; GE Healthcare; IXICO Ltd.; Janssen Alzheimer Immunotherapy Research \& Development, LLC.; Johnson \& Johnson Pharmaceutical Research \& Development LLC.; Lumosity; Lundbeck; Merck \& Co., Inc.;
Meso Scale Diagnostics, LLC.; NeuroRx Research; Neurotrack Technologies; Novartis Pharmaceuticals Corporation; Pfizer Inc.; Piramal Imaging; Servier; Takeda Pharmaceutical Company; and Transition Therapeutics. The Canadian Institutes of Health Research is providing funds to support ADNI clinical sites in Canada. Private sector contributions are facilitated by the Foundation for the National Institutes of Health (\url{www.fnih.org}). The grantee organization is the Northern California Institute for Research and Education, and the study is coordinated by the Alzheimer's Therapeutic Research Institute at the University of Southern California. ADNI data are disseminated by the Laboratory for Neuro Imaging at the University of Southern California.

%\bibliographystyle{model2-names.bst}\ %% biboptions{authoryear}\
%\bibliography{references}

\begin{thebibliography}{23}
\expandafter\ifx\csname natexlab\endcsname\relax\def\natexlab#1{#1}\fi
\providecommand{\url}[1]{\texttt{#1}}
\providecommand{\href}[2]{#2}
\providecommand{\path}[1]{#1}
\providecommand{\DOIprefix}{doi:}
\providecommand{\ArXivprefix}{arXiv:}
\providecommand{\URLprefix}{URL: }
\providecommand{\Pubmedprefix}{pmid:}
\providecommand{\doi}[1]{\href{http://dx.doi.org/#1}{\path{#1}}}
\providecommand{\Pubmed}[1]{\href{pmid:#1}{\path{#1}}}
\providecommand{\bibinfo}[2]{#2}
\ifx\xfnm\relax \def\xfnm[#1]{\unskip,\space#1}\fi
%Type = Article
\bibitem[{Abadi et~al.(2016)Abadi, Agarwal, Barham, Brevdo, Chen, Citro,
  Corrado, Davis, Dean, Devin et~al.}]{Abadi2016}
\bibinfo{author}{Abadi, M.}, \bibinfo{author}{Agarwal, A.},
  \bibinfo{author}{Barham, P.}, \bibinfo{author}{Brevdo, E.},
  \bibinfo{author}{Chen, Z.}, \bibinfo{author}{Citro, C.},
  \bibinfo{author}{Corrado, G.S.}, \bibinfo{author}{Davis, A.},
  \bibinfo{author}{Dean, J.}, \bibinfo{author}{Devin, M.}, et~al.,
  \bibinfo{year}{2016}.
\newblock \bibinfo{title}{Tensorflow: Large-scale machine learning on
  heterogeneous distributed systems}.
\newblock \bibinfo{journal}{arXiv preprint arXiv:1603.04467} .
%Type = Article
\bibitem[{Arabi et~al.(2021)Arabi, AkhavanAllaf, Sanaat, Shiri and
  Zaidi}]{Arabi2021}
\bibinfo{author}{Arabi, H.}, \bibinfo{author}{AkhavanAllaf, A.},
  \bibinfo{author}{Sanaat, A.}, \bibinfo{author}{Shiri, I.},
  \bibinfo{author}{Zaidi, H.}, \bibinfo{year}{2021}.
\newblock \bibinfo{title}{The promise of artificial intelligence and deep
  learning in {PET} and {SPECT} imaging}.
\newblock \bibinfo{journal}{Physica Medica} \bibinfo{volume}{83},
  \bibinfo{pages}{122--137}.
%Type = Article
\bibitem[{Bullich et~al.(2017)Bullich, Seibyl, Catafau, Jovalekic, Koglin,
  Barthel, Sabri and De~Santi}]{Bullich2017}
\bibinfo{author}{Bullich, S.}, \bibinfo{author}{Seibyl, J.},
  \bibinfo{author}{Catafau, A.M.}, \bibinfo{author}{Jovalekic, A.},
  \bibinfo{author}{Koglin, N.}, \bibinfo{author}{Barthel, H.},
  \bibinfo{author}{Sabri, O.}, \bibinfo{author}{De~Santi, S.},
  \bibinfo{year}{2017}.
\newblock \bibinfo{title}{Optimized classification of {18F}-{Florbetaben} {PET}
  scans as positive and negative using an {SUVR} quantitative approach and
  comparison to visual assessment}.
\newblock \bibinfo{journal}{NeuroImage: Clinical} \bibinfo{volume}{15},
  \bibinfo{pages}{325--332}.
%Type = Article
\bibitem[{Chen et~al.(2019)Chen, Gong, de~Carvalho~Macruz, Xu, Boumis,
  Khalighi, Poston, Sha, Greicius, Mormino et~al.}]{Chen2019}
\bibinfo{author}{Chen, K.T.}, \bibinfo{author}{Gong, E.},
  \bibinfo{author}{de~Carvalho~Macruz, F.B.}, \bibinfo{author}{Xu, J.},
  \bibinfo{author}{Boumis, A.}, \bibinfo{author}{Khalighi, M.},
  \bibinfo{author}{Poston, K.L.}, \bibinfo{author}{Sha, S.J.},
  \bibinfo{author}{Greicius, M.D.}, \bibinfo{author}{Mormino, E.}, et~al.,
  \bibinfo{year}{2019}.
\newblock \bibinfo{title}{Ultra--low-dose 18f-florbetaben amyloid {PET} imaging
  using deep learning with multi-contrast {MRI} inputs}.
\newblock \bibinfo{journal}{Radiology} \bibinfo{volume}{290},
  \bibinfo{pages}{649--656}.
%Type = Article
\bibitem[{Gong et~al.(2018)Gong, Guan, Liu and Qi}]{Gong2018}
\bibinfo{author}{Gong, K.}, \bibinfo{author}{Guan, J.}, \bibinfo{author}{Liu,
  C.C.}, \bibinfo{author}{Qi, J.}, \bibinfo{year}{2018}.
\newblock \bibinfo{title}{Pet image denoising using a deep neural network
  through fine tuning}.
\newblock \bibinfo{journal}{IEEE Transactions on Radiation and Plasma Medical
  Sciences} \bibinfo{volume}{3}, \bibinfo{pages}{153--161}.
%Type = Inproceedings
\bibitem[{Huang and Belongie(2017)}]{Huang2017}
\bibinfo{author}{Huang, X.}, \bibinfo{author}{Belongie, S.},
  \bibinfo{year}{2017}.
\newblock \bibinfo{title}{Arbitrary style transfer in real-time with adaptive
  instance normalization}, in: \bibinfo{booktitle}{Proceedings of the IEEE
  international conference on computer vision}, pp.
  \bibinfo{pages}{1501--1510}.
%Type = Article
\bibitem[{Jeong et~al.(2021)Jeong, Park, Jeong, Yoon, Jeon, Cho and
  Kang}]{Jeong2021}
\bibinfo{author}{Jeong, Y.J.}, \bibinfo{author}{Park, H.S.},
  \bibinfo{author}{Jeong, J.E.}, \bibinfo{author}{Yoon, H.J.},
  \bibinfo{author}{Jeon, K.}, \bibinfo{author}{Cho, K.}, \bibinfo{author}{Kang,
  D.Y.}, \bibinfo{year}{2021}.
\newblock \bibinfo{title}{Restoration of amyloid {PET} images obtained with
  short-time data using a generative adversarial networks framework}.
\newblock \bibinfo{journal}{Scientific reports} \bibinfo{volume}{11},
  \bibinfo{pages}{1--11}.
%Type = Article
\bibitem[{Jeong et~al.(2017)Jeong, Yoon and Kang}]{Jeong2017}
\bibinfo{author}{Jeong, Y.J.}, \bibinfo{author}{Yoon, H.J.},
  \bibinfo{author}{Kang, D.Y.}, \bibinfo{year}{2017}.
\newblock \bibinfo{title}{Assessment of change in glucose metabolism in white
  matter of amyloid-positive patients with alzheimer disease using {F-18 FDG
  PET}}.
\newblock \bibinfo{journal}{Medicine} \bibinfo{volume}{96}.
%Type = Inproceedings
\bibitem[{Joshi et~al.(2012)Joshi, Dredze, Cohen and Rose}]{Joshi2012}
\bibinfo{author}{Joshi, M.}, \bibinfo{author}{Dredze, M.},
  \bibinfo{author}{Cohen, W.}, \bibinfo{author}{Rose, C.},
  \bibinfo{year}{2012}.
\newblock \bibinfo{title}{Multi-domain learning: when do domains matter?}, in:
  \bibinfo{booktitle}{Proceedings of the 2012 Joint Conference on Empirical
  Methods in Natural Language Processing and Computational Natural Language
  Learning}, pp. \bibinfo{pages}{1302--1312}.
%Type = Article
\bibitem[{Kingma and Ba(2014)}]{Kingma2014}
\bibinfo{author}{Kingma, D.P.}, \bibinfo{author}{Ba, J.}, \bibinfo{year}{2014}.
\newblock \bibinfo{title}{Adam: A method for stochastic optimization}.
\newblock \bibinfo{journal}{arXiv preprint arXiv:1412.6980} .
%Type = Inproceedings
\bibitem[{Mao et~al.(2017)Mao, Li, Xie, Lau, Wang and Paul~Smolley}]{Mao2017}
\bibinfo{author}{Mao, X.}, \bibinfo{author}{Li, Q.}, \bibinfo{author}{Xie, H.},
  \bibinfo{author}{Lau, R.Y.}, \bibinfo{author}{Wang, Z.},
  \bibinfo{author}{Paul~Smolley, S.}, \bibinfo{year}{2017}.
\newblock \bibinfo{title}{Least squares generative adversarial networks}, in:
  \bibinfo{booktitle}{Proceedings of the IEEE international conference on
  computer vision}, pp. \bibinfo{pages}{2794--2802}.
%Type = Inproceedings
\bibitem[{Nam and Han(2016)}]{Nam2016}
\bibinfo{author}{Nam, H.}, \bibinfo{author}{Han, B.}, \bibinfo{year}{2016}.
\newblock \bibinfo{title}{Learning multi-domain convolutional neural networks
  for visual tracking}, in: \bibinfo{booktitle}{Proceedings of the IEEE
  conference on computer vision and pattern recognition}, pp.
  \bibinfo{pages}{4293--4302}.
%Type = Article
\bibitem[{Ouyang et~al.(2019)Ouyang, Chen, Gong, Pauly and
  Zaharchuk}]{Ouyang2019}
\bibinfo{author}{Ouyang, J.}, \bibinfo{author}{Chen, K.T.},
  \bibinfo{author}{Gong, E.}, \bibinfo{author}{Pauly, J.},
  \bibinfo{author}{Zaharchuk, G.}, \bibinfo{year}{2019}.
\newblock \bibinfo{title}{Ultra-low-dose pet reconstruction using generative
  adversarial network with feature matching and task-specific perceptual loss}.
\newblock \bibinfo{journal}{Medical physics} \bibinfo{volume}{46},
  \bibinfo{pages}{3555--3564}.
%Type = Article
\bibitem[{Sanaat et~al.(2021)Sanaat, Shiri, Arabi, Mainta, Nkoulou and
  Zaidi}]{Sanaat2021}
\bibinfo{author}{Sanaat, A.}, \bibinfo{author}{Shiri, I.},
  \bibinfo{author}{Arabi, H.}, \bibinfo{author}{Mainta, I.},
  \bibinfo{author}{Nkoulou, R.}, \bibinfo{author}{Zaidi, H.},
  \bibinfo{year}{2021}.
\newblock \bibinfo{title}{Deep learning-assisted ultra-fast/low-dose whole-body
  {PET}/{CT} imaging}.
\newblock \bibinfo{journal}{European journal of nuclear medicine and molecular
  imaging} \bibinfo{volume}{48}, \bibinfo{pages}{2405--2415}.
%Type = Article
\bibitem[{Sergeev and Balso(2018)}]{horovod2018}
\bibinfo{author}{Sergeev, A.}, \bibinfo{author}{Balso, M.D.},
  \bibinfo{year}{2018}.
\newblock \bibinfo{title}{Horovod: fast and easy distributed deep learning in
  {TensorFlow}}.
\newblock \bibinfo{journal}{arXiv preprint arXiv:1802.05799} .
%Type = Article
\bibitem[{Wang et~al.(2020)Wang, Ye and De~Man}]{Wang2020}
\bibinfo{author}{Wang, G.}, \bibinfo{author}{Ye, J.C.},
  \bibinfo{author}{De~Man, B.}, \bibinfo{year}{2020}.
\newblock \bibinfo{title}{Deep learning for tomographic image reconstruction}.
\newblock \bibinfo{journal}{Nature Machine Intelligence} \bibinfo{volume}{2},
  \bibinfo{pages}{737--748}.
%Type = Article
\bibitem[{Wang et~al.(2021)Wang, Baratto, Hawk, Theruvath, Pribnow, Thakor,
  Gatidis, Lu, Gummidipundi, Garcia-Diaz et~al.}]{Wang2021}
\bibinfo{author}{Wang, Y.R.}, \bibinfo{author}{Baratto, L.},
  \bibinfo{author}{Hawk, K.E.}, \bibinfo{author}{Theruvath, A.J.},
  \bibinfo{author}{Pribnow, A.}, \bibinfo{author}{Thakor, A.S.},
  \bibinfo{author}{Gatidis, S.}, \bibinfo{author}{Lu, R.},
  \bibinfo{author}{Gummidipundi, S.E.}, \bibinfo{author}{Garcia-Diaz, J.},
  et~al., \bibinfo{year}{2021}.
\newblock \bibinfo{title}{Artificial intelligence enables whole-body positron
  emission tomography scans with minimal radiation exposure}.
\newblock \bibinfo{journal}{European journal of nuclear medicine and molecular
  imaging} \bibinfo{volume}{48}, \bibinfo{pages}{2771--2781}.
%Type = Article
\bibitem[{Xiang et~al.(2017)Xiang, Qiao, Nie, An, Lin, Wang and
  Shen}]{Xiang2017}
\bibinfo{author}{Xiang, L.}, \bibinfo{author}{Qiao, Y.}, \bibinfo{author}{Nie,
  D.}, \bibinfo{author}{An, L.}, \bibinfo{author}{Lin, W.},
  \bibinfo{author}{Wang, Q.}, \bibinfo{author}{Shen, D.}, \bibinfo{year}{2017}.
\newblock \bibinfo{title}{Deep auto-context convolutional neural networks for
  standard-dose {PET} image estimation from low-dose {PET}/{MRI}}.
\newblock \bibinfo{journal}{Neurocomputing} \bibinfo{volume}{267},
  \bibinfo{pages}{406--416}.
%Type = Article
\bibitem[{Xue et~al.(2021)Xue, Zhang, Zou, Zhang, Zhou, Tie, Wan, Teng, Li,
  Liang et~al.}]{Xue2021}
\bibinfo{author}{Xue, H.}, \bibinfo{author}{Zhang, Q.}, \bibinfo{author}{Zou,
  S.}, \bibinfo{author}{Zhang, W.}, \bibinfo{author}{Zhou, C.},
  \bibinfo{author}{Tie, C.}, \bibinfo{author}{Wan, Q.}, \bibinfo{author}{Teng,
  Y.}, \bibinfo{author}{Li, Y.}, \bibinfo{author}{Liang, D.}, et~al.,
  \bibinfo{year}{2021}.
\newblock \bibinfo{title}{{LCPR}-{Net}: low-count {PET} image reconstruction
  using the domain transform and cycle-consistent generative adversarial
  networks}.
\newblock \bibinfo{journal}{Quantitative imaging in medicine and surgery}
  \bibinfo{volume}{11}, \bibinfo{pages}{749}.
%Type = Article
\bibitem[{Yang et~al.(2018)Yang, Yan, Zhang, Yu, Shi, Mou, Kalra, Zhang, Sun
  and Wang}]{Yang2018}
\bibinfo{author}{Yang, Q.}, \bibinfo{author}{Yan, P.}, \bibinfo{author}{Zhang,
  Y.}, \bibinfo{author}{Yu, H.}, \bibinfo{author}{Shi, Y.},
  \bibinfo{author}{Mou, X.}, \bibinfo{author}{Kalra, M.K.},
  \bibinfo{author}{Zhang, Y.}, \bibinfo{author}{Sun, L.},
  \bibinfo{author}{Wang, G.}, \bibinfo{year}{2018}.
\newblock \bibinfo{title}{Low-dose {CT} image denoising using a generative
  adversarial network with {Wasserstein} distance and perceptual loss}.
\newblock \bibinfo{journal}{IEEE transactions on medical imaging}
  \bibinfo{volume}{37}, \bibinfo{pages}{1348--1357}.
%Type = Article
\bibitem[{Ye et~al.(2018)Ye, Han and Cha}]{Ye2018}
\bibinfo{author}{Ye, J.C.}, \bibinfo{author}{Han, Y.}, \bibinfo{author}{Cha,
  E.}, \bibinfo{year}{2018}.
\newblock \bibinfo{title}{Deep convolutional framelets: A general deep learning
  framework for inverse problems}.
\newblock \bibinfo{journal}{SIAM Journal on Imaging Sciences}
  \bibinfo{volume}{11}, \bibinfo{pages}{991--1048}.
%Type = Article
\bibitem[{Zhou et~al.(2020)Zhou, Schaefferkoetter, Tham, Huang and
  Yan}]{Zhou2020}
\bibinfo{author}{Zhou, L.}, \bibinfo{author}{Schaefferkoetter, J.D.},
  \bibinfo{author}{Tham, I.W.}, \bibinfo{author}{Huang, G.},
  \bibinfo{author}{Yan, J.}, \bibinfo{year}{2020}.
\newblock \bibinfo{title}{Supervised learning with cyclegan for low-dose {FDG
  PET} image denoising}.
\newblock \bibinfo{journal}{Medical image analysis} \bibinfo{volume}{65},
  \bibinfo{pages}{101770}.
%Type = Inproceedings
\bibitem[{Zhou et~al.(2021)Zhou, Greenspan, Davatzikos, Duncan, Van~Ginneken,
  Madabhushi, Prince, Rueckert and Summers}]{Zhou2021}
\bibinfo{author}{Zhou, S.K.}, \bibinfo{author}{Greenspan, H.},
  \bibinfo{author}{Davatzikos, C.}, \bibinfo{author}{Duncan, J.S.},
  \bibinfo{author}{Van~Ginneken, B.}, \bibinfo{author}{Madabhushi, A.},
  \bibinfo{author}{Prince, J.L.}, \bibinfo{author}{Rueckert, D.},
  \bibinfo{author}{Summers, R.M.}, \bibinfo{year}{2021}.
\newblock \bibinfo{title}{A review of deep learning in medical imaging: Imaging
  traits, technology trends, case studies with progress highlights, and future
  promises}, in: \bibinfo{booktitle}{Proceedings of the IEEE}, pp.
  \bibinfo{pages}{820--838}.
\end{thebibliography}

\end{document}